\documentclass[twocolumn,amssymb,nobibnotes,showpacs,superscriptaddress,aps,pra,longbibliography]{revtex4-1}

\usepackage{amsmath,graphicx}
\usepackage[colorlinks,linkcolor=blue,urlcolor=blue,anchorcolor=blue,citecolor=blue]{hyperref}

\begin{document}
	
\title{Interfering pathways for photon blockade in cavity QED with one and two qubits}
	
\author{K. Hou}
\affiliation{MOE Key Laboratory of Advanced Micro-Structured Materials, School of Physics Science and Engineering,Tongji University, Shanghai, China 200092}
\affiliation{Department of Mathematics and Physics, Anhui JianZhu University, Hefei 230601,China}
\author{C. J. Zhu}
\email[Corresponding author:]{cjzhu@tongji.edu.cn}
\affiliation{MOE Key Laboratory of Advanced Micro-Structured Materials, School of Physics Science and Engineering,Tongji University, Shanghai, China 200092}
\author{Y. P. Yang}
\email[Corresponding author:]{yang\_yaping@tongji.edu.cn}
\affiliation{MOE Key Laboratory of Advanced Micro-Structured Materials, School of Physics Science and Engineering,Tongji University, Shanghai, China 200092}
\author{G. S. Agarwal}
\email[Corresponding author:]{girish.agarwal@tamu.edu}
\affiliation{Institute for Quantum Science and Engineering, and Department of Biological and Agricultural Engineering Texas A$\&$M University,
College Station, Texas 77843, USA}

\begin{abstract}
We theoretically study the quantum interference induced photon blockade phenomenon in atom cavity QED system, where the destructive interference between two different transition pathways prohibits the two-photon excitation. Here, we first explore the single atom cavity QED system via an atom or cavity drive. We show that the cavity-driven case will lead to the quantum interference induced photon blockade under a specific condition, but the atom driven case can't result in such interference induced photon blockade. Then, we investigate the two atoms case, and find that an additional transition pathway appears in the atom-driven case. We show that this additional transition pathway results in the quantum interference induced photon blockade only if the atomic resonant frequency is different from the cavity mode frequency. Moreover, in this case, the condition for realizing the interference induced photon blockade is independent of the system's intrinsic parameters, which can be used to generate antibunched photon source both in weak and strong coupling regimes.
\end{abstract}
	
\maketitle
	
\section{Introduction}
The phenomenon of quantum interference (QI) effect~\cite{ficek2005quantum} occurs between different photon transmission pathways, leading to the inhibition/attenuation of absorption by the destructive interference~\cite{agarwal2012quantum}. With its fascinating physical mechanism, many novel quantum effects and corresponding applications have emerged, e.g., electromagnetically induced transparency (EIT)~\cite{EIT2005}, coherent population trapping (CPT)~\cite{CPT2017}, laser without inversion~\cite{LWI1989}, light with ultraslow group velocity~\cite{baba2008slow} and so on. Specifically, quantum interference between two photons has been observed in experiments~\cite{beugnon2006quantum,araneda2018interference}. Moreover, the QI effect can result in a new type of photon blockade (PB) phenomenon in cavity QED systems~\cite{PhysRevA.83.021802}.
%However,  with the development of quantum information and quantum communication, it is not only necessary to make use of this interference effect as a resource for quantum data processing, but also to generate entanglement in a quantum network which strongly relied on encoding quantum information in single photons~\cite{RevModPhys.87.1379}.

The traditional PB results from the anharmonicity of the Jaynes Cummings ladder in atom cavity QED systems~\cite{PhysRevLett.79.1467}, where the absorption of the first photon blocks the transmission of the second photon. As a result, one can observe an orderly output of photons one by one with strong photon antibunching and the sub-Poissonian statistics. Up to now, the traditional PB has been experimentally demonstrated in various quantum systems, including atoms~\cite{birnbaum2005photon,dayan2008photon}, quantum dot~\cite{faraon2008coherent} and ions cavity quantum electrodynamics systems~\cite{PhysRevLett.120.073001} as well as the circuit-QED systems~\cite{PhysRevLett.106.243601,PhysRevLett.107.053602}.

To blockade the transmission of the second photon, strong coupling limit is required in the traditional PB, which is challenging in experiments, especially for semiconductor cavity QED systems. In addition, with current experimental techniques, the second order correlation function $g^{(2)}$, which is a signature for PB, can't be close to zero to achieve perfect antibunched photons. This is because the two-photon and multiphoton excitations cannot be inhibited due to the energy broadening. In the light of these disadvantages, a novel physical mechanism based on quantum interference is proposed to generate strong PB phenomenon~\cite{PhysRevLett.104.183601}. In literature, this quantum interference induced PB is known as the unconventional photon blockade (UPB). In general, there exist two methods to open an additional transition pathway for generating the quantum interference effect. One is by adding an auxiliary cavity with the coherent mode coupling~\cite{PhysRevLett.104.183601,Ferretti_2013,PhysRevA.90.033809}, while the other is by adding an auxiliary driving field ~\cite{tang2015quantum,Flayac2017UPB}.
%the shows that strong antibunching  can be obtained in two coupled photonic cavities with weak Kerr nonlinearity. 

%This counterintuitive effect can be interpreted in terms of quantum interference between different excitation pathways or optimal squeezing\cite{Flayac2017UPB}. 

%To date, the UPB has been theoretically investigated in various systems, such as {\bf Hou: rewrite this para.} QD-bimodal nanocavity cavity\cite{PhysRevLett.108.183601, PhysRevA.89.043832}, four-wave mixing in a three mode system with weak Kerr-type nonlinearity\cite{PhysRevA.90.063805}, weakly nonlinear passive materials coupled photonic crystal resonators\cite{ferretti2013optimal}, photonic molecule consisting of two driven nonlinear cavity modes\cite{PhysRevA.90.043822}, quantum well coupled to the icropillar optical cavity\cite{PhysRevA.90.033807}, two cavities coupled to $\chi^{(2)}$ nonlinearity\cite{PhysRevA.92.023838, PhysRevA.89.031803} or $\chi^{(3)}$ nonlinearity\cite{Shen:15}, solid-state cavity QED system\cite{PhysRevB.96.201201}, two-emitter-cavity systems\cite{PhysRevA.96.011801}, optomechanical systems\cite{PhysRevA.92.033806,PhysRevA.98.013826}, optical parametric amplifie coupled to cavity system\cite{PhysRevA.96.053827}, as well as the hybrid quantum plasmonic system\cite{PhysRevA.98.033834}. 

To date, the UPB has been theoretically investigated in various systems, such as two tunnel-coupled cavity system~\cite{PhysRevA.90.033807,PhysRevA.90.043822,PhysRevA.92.023838,PhysRevA.89.031803,Shen:15}, quantum dots~\cite{PhysRevLett.108.183601,PhysRevA.89.043832,PhysRevB.96.201201}, quantum well~\cite{PhysRevA.90.033807}, nanomechanical resonator~\cite{PhysRevA.94.063853}, optomechanical system~\cite{PhysRevA.92.033806,PhysRevA.98.013826}, optical parametric amplifier system~\cite{PhysRevA.96.053827}, as well as the hybrid quantum plasmonic system~\cite{PhysRevA.98.033834}. Recently, the UPB is experimentally demonstrated in two coupled superconducting resonators~\cite{PhysRevLett.121.043602} and a single quantum dot cavity QED system driven by two orthogonally polarized modes~\cite{PhysRevLett.121.043601}.

%, which opens the door to generate the non-classical light in weak single photon nonlinearity regime. As far as we know, the excitation of the system in the UPB mentioned here all via the driving resonator. However, Rempe et al\cite{hamsen2017two} shows that driving the quantum emitter instead of the resonator improves the nonlinear response of the strongly coupled cavity quantum electrodynamic(CQED) system, resulting in both one- and two-photon PB.

In this paper, we theoretically investigate the quantum interference induced photon blockade in atom cavity QED systems via an atom or cavity drive. Using the amplitude method, we obtain the condition for observing the interference induced photon blockade in the atom cavity QED system via a cavity drive, i.e., equation (5). To the best of our knowledge, this condition has not yet been reported in the literature. Opposite to the work in Ref.~\cite{PhysRevA.83.021802}, we show that the interference induced photon blockade can be realized without requiring the weak nonlinearity or auxiliary pumping field. Moreover, we show that additional transition pathways take place by adding another atom in the cavity. These additional transition pathways result in the interference induced photon blockade even via an atom drive. Therefore, extremely strong antibunching photons can be accomplished, leading to the value of the second-order correlation function smaller than unity. 

The paper is arranged as follows. In section II, we first study the single atom case via a cavity drive or an atom drive. We show that the destructive interference can only be observed in the case of the cavity drive because the destructive interference can be achieved between transition pathways with odd number of photons. However, there is no interference in the case of the atom drive. This is contrary to intuition as one generally views that atom drive and cavity drive should be similar physics. In section III, we study the two atoms case also via a cavity drive or an atom drive. We show that the odd-photon transition induced destructive interference in cavity driven case can be improved by the collective coupling enhancement, leading to a significant improvement of the photon blockade phenomenon. In the atom-driven case, surprisingly, we find that an additional transition pathway appears in the presence of the second atom, yielding interference between transitions with even number of photons. If the atomic resonant frequency is the same as the cavity mode frequency, we show that this interference is constructive, so that the photon blockade can not be observed. In section IV, we show that this even-photon interference can be destructive when the atomic resonant frequency is not the same as the cavity mode frequency. As a result, two transition pathways become distinguishable, leading to the interference induced photon blockade. We also show that the condition for realizing this even-photon interference induced PB in atom-driven scheme is insensitive to the atom-cavity coupling strength. Thus, one can increase the coupling strength to obtain reasonable photon number with strong PB effect. 
%we here we theoretically study the possibility of realizing UPB in an atom CQED system for different drive mode. For the atoms coupled to the cavity without detuning, we find that the optimal antibuching corresponding UPB can be observed in resonance driven cavity case instead of resonance driven atoms system. Nevertheless, the atoms-driven CQED system also exhibits optimal UPB by regulating the corresponding detunings. We show that this strong antibunching effect is the result of the destructive quantum interference between the time-order paths\cite{PhysRevLett.93.093002}. The optimal condition independent of the intrinsic system parameters is obtained by analytical conclusion in this case. Hence, it relax the requirement for fine interrelations between emitter and resonator in general UPB. As a result, the photon correlation of the output field is not sensitive to the coupling strength when the optimal condition satisfied, which allows us to observe the UPB both in the weak- and strong-coupling regime. Therefore, one can enlarge the mean photon number for strong photon antibunching with a moderate atom-cavity coupling. The validation of our results is evidenced by numerical simulations.

\section{single atom cavity QED system}
\begin{figure}[htbp]
	\centering
	\includegraphics[width=\linewidth]{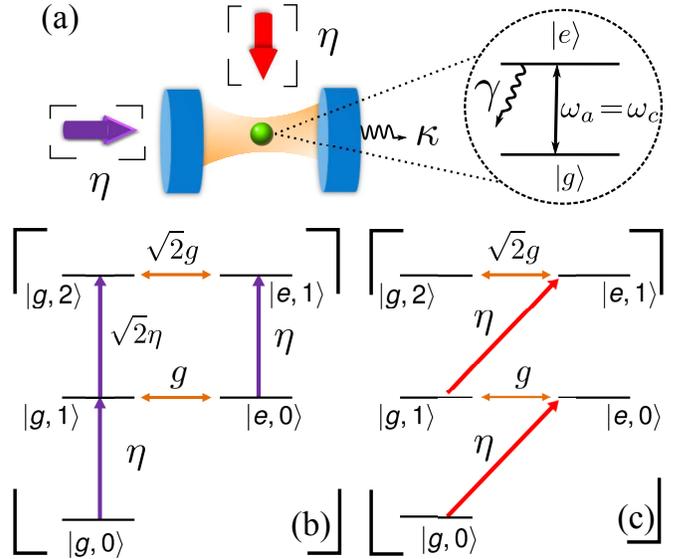}
	\caption{(Color online) (a) Sketch of a two-level atom with resonant frequency $\omega_a$ trapped in a single-mode cavity. The red (purple) arrow corresponds to the atom (cavity) drive with driving strength $\eta$. $|g\rangle$ ($|e\rangle$) is the ground (excited) state of the atom. $\gamma$ and $\kappa$ are the atomic and cavity decay rates, respectively. Panels (b) and (c) show the transition pathways in an atom- and cavity-driven cases, respectively. Here yellow arrows represent the atom-cavity coupling with strength $g$. $|\alpha,n\rangle\ (\alpha=g,e)$ is the product state of atomic state $|\alpha\rangle$ and photon state $|n\rangle$.}~\label{fig:fig1}
\end{figure}
First, we consider a typical single atom cavity QED system as shown in Fig.~\ref{fig:fig1}(a). In the frame rotating at the driving frequency $\omega_d$, the system Hamiltonian is written as~\cite{hamsen2017two} (setting $\hbar=1$)
\begin{equation}\label{eq:H1}
H_{1}=-\Delta_ca^\dagger a-\Delta_a\sigma^\dagger\sigma+g(\sigma a^\dagger+\sigma^\dagger a)+H_d,
\end{equation}
where $H_d$ is the driving term. If the cavity (atom) is driven by a coherent field, $H_d=\eta(a+a^\dagger)$ [$H_d=\eta(\sigma_j+\sigma_j^\dagger)$] with $\eta$ being the driving strength. Here, $\Delta_c=\omega_d-\omega_{c}$ and $\Delta_a=\omega_d-\omega_{\rm a}$ are the detunings for the cavity and atom, respectively. $a$ ($a^\dagger$) is the annihilation (creation) operator of the cavity mode with the resonant frequency $\omega_c$. $\sigma$ ($\sigma^\dagger$) denotes the lowering (raising) operator of the two-level atom with the resonant transition frequency $\omega_a$. $g$ is the atom-cavity coupling strength. 
%For all calculations in this section, we consider the driving field is resonant with both of the cavity and atom, i.e., $\Delta_a=\Delta_c=0$.

The dynamics of this open quantum system is governed by the master equation, i.e., 
\begin{eqnarray}\label{eq:master}
\frac{\partial \rho}{\partial t}=-i[H,\rho]
+\mathcal{L}_{\kappa} \rho+\mathcal{L}_{\gamma} \rho,
\end{eqnarray}
where $\mathcal{L}_{\kappa} \rho=\kappa(2a\rho a^{\dagger}-a^{\dagger}a\rho-\rho a^{\dagger} a)$ and $\mathcal{L}_{\gamma} \rho=\gamma(2\sigma\rho\sigma^{\dagger}-\sigma^{\dagger}\sigma\rho-\rho\sigma^{\dagger}\sigma)$ describe the dissipations of the cavity and atom with the decay rate $\kappa$ and $\gamma$, respectively. Numerically solving Eq.~(\ref{eq:master}), one can obtain the second-order photon-photon correlation function $g^{(2)}(0)=\langle a^\dagger a^\dagger aa\rangle/\langle a^\dagger a\rangle^2$ in the steady state. In general, the value of $g^{(2)}(0)$ characterizes the probability of detecting two photons at the same time. If $g^{(2)}(0)>1$, two photons will be detected simultaneously. However, if $g^{(2)}(0)<1$, the output of photons has antibunching behavior, i.e., detecting photons one by one. In the following, we will discuss this phenomenon in detail.

\vskip 5pt
{\it Cavity-driven scheme.} - We consider that the cavity is driven by a coherent field and assume $\Delta_a=\Delta_c=0$ for simplicity. Thus, the transition pathways of this cavity-driven scheme is shown in Fig.~\ref{fig:fig1}(b), where $|\alpha,n\rangle$ represent the states in the atom-cavity product space. Obviously, there exist two transition pathways corresponding to the transition from $|g,1\rangle$ to $|g,2\rangle$. The first one is $|g,1\rangle\overset{\sqrt{2}\eta}{\longrightarrow}|g,2\rangle$ transition, and the second one is $|g,1\rangle\overset{g}{\longleftrightarrow}|e,0\rangle\overset{\eta}{\longrightarrow}|e,1\rangle\overset{\sqrt{2}g}{\longleftrightarrow}|g,2\rangle$ transition. If the destructive interference takes place between these two transition pathways, the two-photon excitation will be not allowed and the probability for detecting the state $|g,2\rangle$ will be zero. Then, one can achieve quantum interference induced photon blockade phenomenon, leading to an output of antibunched photons.

To obtain the condition for this destructive interference, we assume the system wavefunction $\Psi\approx\sum_{n=0}^2C_{g,n}|g,n\rangle+\sum_{n=0}^1C_{e,n}|e,n\rangle$, where $|C_{\alpha,n}|^2$ ($\alpha=g,e$) is the probability for detecting the state $|\alpha,n\rangle$. Here, other states in larger photon-number spaces  have been neglected since the driving field is not strong enough to excite these states. Then, the dynamical equations for the amplitudes of each state can be written as 
\begin{subequations}\label{eq:c1}
\begin{eqnarray}
&&i\dot{C}_{g,1}=gC_{e,0}-i\frac{\kappa}{2} C_{g,1}+\eta C_{g,0}+\sqrt{2}\eta C_{g,2},\\
&&i\dot{C}_{g,2}=\sqrt{2}gC_{e,1}-i\kappa C_{g,2}+\sqrt{2}\eta C_{g,1},\\
&&i\dot{C}_{e,0}=gC_{g,1}-i\frac{\gamma}{2}C_{e,0}+\sqrt{2}\eta C_{e,1},\\
&&i\dot{C}_{e,1}=\sqrt{2}gC_{g,2}-i\left(\frac{\kappa+\gamma}{2}\right)C_{e,1}+\eta C_{e,0}.
\end{eqnarray}
\end{subequations}

Using the perturbation method~\cite{PhysRevA.82.013841, PhysRevA.83.021802} and solving above equations under the steady state approximation, one can obtain 
\begin{equation}\label{eq:cg2}
C_{g,2}=\frac{2 \sqrt{2} \eta^2 \left(-\gamma ^2-\gamma  \kappa +4 g^2-4\eta^2\right)}{\left(\gamma  \kappa +4 g^2\right) \left(\gamma  \kappa +4 g^2+\kappa ^2\right)+4\eta^2X},
\end{equation}
with $X=4\eta^2-8g^2+\gamma^2+\kappa^2+\gamma\kappa$. Clearly, the optimal condition for $C_{g,2}=0$ is 
\begin{equation}\label{eq:one atom optimal g20}
g=\frac{1}{2}\sqrt{\gamma^2+\gamma\kappa+4\eta^2}.
\end{equation}
Under the weak driving condition, the second-order correlation function can be expressed as $g^{(2)}(0)\approx2|C_{g,2}|^2/|C_{g,1}|^4$, yielding  $g^{(2)}(0)\rightarrow 0$ if Eq.~(\ref{eq:one atom optimal g20}) is satisfied.

It is worth to point out that Eq.~(\ref{eq:one atom optimal g20}) is the condition for achieving destructive interference induced photon blockade in a single atom cavity QED system, as far as we known, which has not yet been reported before. Compared with the works in Refs.~\cite{PhysRevA.82.013841,PhysRevLett.121.043602,PhysRevLett.121.043601}, our system is much simpler than their proposals, where the destructive interference results from the transition pathways between two cavity modes and weak nonlinearity is essential to realize the destructive interference. Eq.~(\ref{eq:one atom optimal g20}) implies that, under weak atom cavity coupling, the photon blockade can also be accomplished in a typical single atom cavity QED system via the destructive interference. It is noted that Eq.~(\ref{eq:one atom optimal g20}) is only valid for the weak driving field since high-order photon states are assumed to be unexcited. For strong driving field, however, Eq.~(\ref{eq:one atom optimal g20}) is invalid and one can not observe the interference induced photon blockade since high-order photon states will be excited, yielding $g^{(2)}$ larger than unity. 

%So if the single atom CQED system resonantly driving via the cavity, the coupling strength $g$ for optimal UPB is determined by the decay rates $\kappa$ and $\gamma$. The optimal condition is similar to that of a nanomechanical resonator coupled to a qubit\cite{PhysRevA.94.063853}.

%
\begin{figure}[htbp]
	\centering
	\includegraphics[width=\linewidth]{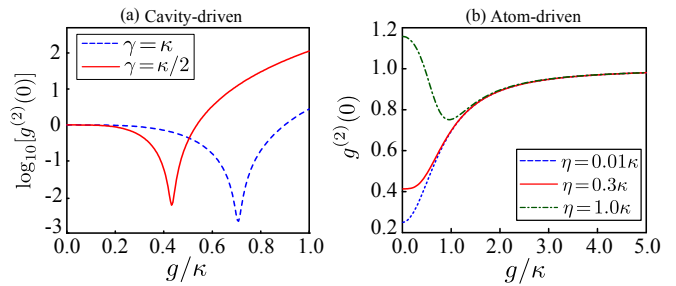}
	\caption{(Color online) The equal-time second-order correlation function $g^{(2)}(0)$ versus the normalized coupling strength $g/\kappa$ in the (a) cavity-driven and (b) atom-driven systems, respectively. In panel (a), we choose $\gamma=\kappa$ (blue dashed curve) and $\gamma=\kappa/2$ (red solid curve), respectively. The driving strength $\eta=0.01\kappa$. In panel (b), we choose $\gamma=\kappa$, but the driving strength is chosen as $\eta=0.01\kappa$ (blue dashed curve), $\eta=0.3\kappa$ (red solid curve) and $\eta=\kappa$ (green dash dotted curve), respectively.}~\label{fig:fig2}
\end{figure}
To verify the above analysis, we numerically solve Eq.~(\ref{eq:master}). In Fig.~\ref{fig:fig2}(a), we plot the equal-time second-order correlation function $g^{(2)}(0)$ as a function of atom-cavity coupling strength with atomic decay rate $\gamma=\kappa$ (blue dashed curve) and $\gamma=\kappa/2$ (red solid curve), respectively. Other system parameters are given by $\Delta_a=\Delta_c=0$ and $\eta=0.01\kappa$. It is clear to see that there exist a minimum in the second-order correlation function at $g=\sqrt{\gamma(\gamma+\kappa)+4\eta^2}/2$ due to the destructive interfere effect. Moreover, the smaller the atomic decay rate is, the smaller the value of $g^{(2)}(0)$ is, leading to a stronger photon blockade effect as shown in Fig.~\ref{fig:fig2}(a).
%To understand the numerical results, we can derive the condition for optimal antibunching , and the amplitudes of the two photon state $|g,2\rangle$ can be expressed as

\vskip 5pt
{\it No quantum interference for atom-driven scheme.} - Contrary to the cavity-driven case, there exists only one transition pathway for the two-photon excitation, i.e., $|g,0\rangle\overset{\eta}{\rightarrow}|e,0\rangle\overset{g}{\rightarrow}|g,1\rangle\overset{\eta}{\rightarrow}|e,1\rangle\overset{\sqrt{2}g}{\rightarrow}|g,2\rangle$ [see Fig.~\ref{fig:fig1}(c)] if one drive the atom directly. Therefore, the two-photon excitations can't be blockaded by the quantum interference effect. In the case of weak driving strength, e.g., $\eta=0.01\kappa$ (blue dashed curve) and $\eta=0.3\kappa$ (red solid curve), the value of $g^{(2)}(0)$ is smaller than unity for small coupling strengths because the high-order states can't be excited by such weak driving fields. With the increase of the atom-cavity coupling strength, the energy splitting become dominant so that the driving field becomes far off-resonant to all states in the system, yielding  $g^{(2)}(0)\rightarrow1$ as shown in Fig.~\ref{fig:fig2}(b). However, in the case of strong driving strength, e.g., $\eta=\kappa$ (green dash dotted curve), states in two-photon space will be excited and one can obtain $g^{(2)}(0)>1$ for small coupling strength $g$. As $g$ increases, it is clear to see that $g^{(2)}(0)<1$ because the anharmonic energy splitting prevents the states in two-photon space from being excited [see the green curve in Fig.~\ref{fig:fig2}(b)]. Furthering increasing $g$, one can also observe $g^{(2)}(0)\rightarrow1$ since the driving field is in far off-resonant state.
%In addition, the value of $g^{(2)}(0)$ grows monotonously with $g$. Intuitively, this means the weaker the coupling strength the better the single photon behavior of the output. However, the weak driving limit leads to a tiny number of photons in the cavity, and hence of single photons emitted from system  can be confirmed by $g^{(2)}(0)<1$\cite{ferretti2013optimal}.

\section{Two atoms cavity QED system}
\begin{figure}[htbp]
	\centering
	\includegraphics[width=\linewidth]{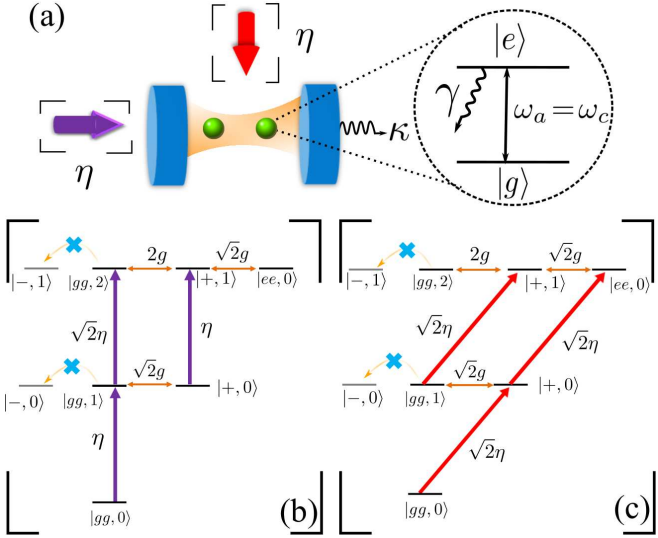}
	\caption{(Color online) (a) Sketch of two two-level atoms trapped in a single-mode cavity. The red (purple) arrow corresponds to the atom  (cavity) drive with driving strength $\eta$. Panels (b) and (c) show the transition pathways in cavity- and atom-driven systems, respectively.}~\label{fig:fig3}
\end{figure}
Next, we consider that two identical atoms are trapped in a single-mode cavity with the same atom-cavity coupling strengths as shown in Fig.~\ref{fig:fig3}(a). The corresponding Hamiltonian is then written as
\begin{equation}\label{eq:H2}
H_{2}=-\Delta_ca^\dagger a+\sum_{j=1}^2[-\Delta_a\sigma_j^\dagger\sigma_j+g(\sigma_j a^\dagger+\sigma_j^\dagger a)]+H_d,
\end{equation}
where the subscript $j$ indicates the $j$-th atom. Likewise, the drive term is given by $H_d=\eta(a^\dag+a)$ for the cavity drive, and  $H_d=\eta\sum_{j=1}^2(\sigma_j+\sigma_j^\dagger)$ for the atom drive, respectively. Then, the master equation describing the dynamics of the system is written as 
\begin{equation}
\frac{\partial\rho}{\partial t}=-i[H_2,\rho]+\mathcal{L}_{\kappa} \rho+\sum_{j=1}^2\mathcal{L}_{\gamma}^{(j)}\rho,
\end{equation}
where $\mathcal{L}^{(j)}_{\gamma} \rho$ represents the dissipation term of the $j$-th atom. 

For mathematical simplicity, we assume $\Delta_a=\Delta_c=0$. In general, such system can be described by using the collective states $\{|gg\rangle, |\pm\rangle, |ee\rangle\}$ as the basis. Here, we define the states  $|\pm\rangle=(|eg\rangle\pm|ge\rangle)/\sqrt{2}$ as the symmetric and anti-symmetric Dicke states, respectively. 

\vskip 5pt
{\it Cavity-driven scheme.} - Since two atoms have the same coupling strength (i.e., in-phase radiation), the anit-symmetric Dicke states $|-,n\rangle$ are dark states, decoupling to other states in this system~\cite{pleinert2017hyperradiance,zhu2017collective}. In the cavity driven scheme, therefore, one can also observe two different transition pathways for the two-photon excitations, corresponding to $|gg,1\rangle\overset{\sqrt{2}\eta}{\rightarrow}|gg,2\rangle$ and  $|gg,1\rangle\overset{\sqrt{2}g}{\rightarrow}|+,0\rangle\overset{\sqrt{2}\eta}{\rightarrow}(|+,1\rangle\overset{\sqrt{2}g}
{\longleftrightarrow}|ee,0\rangle)\overset{2g}{\rightarrow}|gg,2\rangle$, respectively [see Fig.~\ref{fig:fig3}(b)]. 

To obtain the optimal condition for achieving the destructive quantum interference between these two pathways, we also solve the amplitude equations [similar to Eqs.~(\ref{eq:c1})] by assuming the system wavefunction $\Psi\approx\Sigma_{n=0}^2C_{gg,n}|gg,n\rangle+\Sigma_{n=0}^1C_{\pm,n}|\pm,n\rangle+C_{ee,0}|ee,0\rangle$. Under the steady state approximation, the probability amplitude of state $|gg,2\rangle$ is given by
\begin{equation}\label{eq:cgg2}
C_{gg,2}\approx\frac{2 \sqrt{2} \gamma  \eta^2 \left(4 g^2-\gamma ^2-\gamma  \kappa -4\eta^2\right)}{\left(\gamma  \kappa +8 g^2\right) \left(\gamma ^2 \kappa +\gamma  \kappa ^2+8 \gamma  g^2+4 g^2 \kappa \right)}.
\end{equation}
Obviously, one can obtain the same optimal condition for $C_{gg,2}=0$ as given by Eq.~(\ref{eq:one atom optimal g20}), yielding strong photon blockade effect  induced by the destructive interference between these two transition pathways.

\begin{figure}[htbp]
	\centering
	\includegraphics[width=\linewidth]{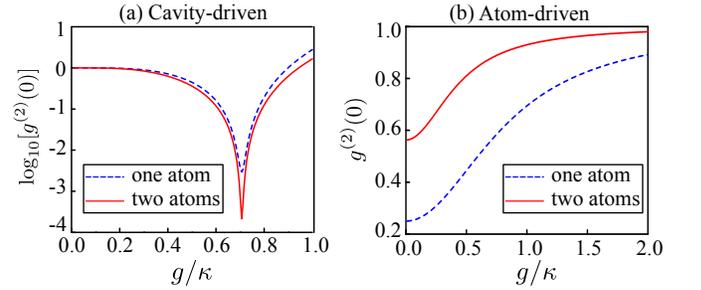}
	\caption{(Color online) The equal-time second-order correlation function $g^{(2)}(0)$ versus the normalized coupling strength $g/\kappa$ in the (a) cavity driven and (b) atom driven systems, respectively. The blue dashed curves correspond to the one atom case, while the red solid curves correspond to the two atoms case. The system parameters are given by $\Delta_a=\Delta_c=0$, $\gamma=\kappa$ and $\eta=0.01\kappa$, respectively.}~\label{fig:fig4}
\end{figure}
In Fig.~\ref{fig:fig4}(a), we plot the equal-time second-order correlation function $g^{(2)}(0)$ as a function of the atom-cavity coupling $g$. The system parameters are given by $\Delta_a=\Delta_c=0$, $\gamma=\kappa$ and $\eta=0.01\kappa$, respectively. It is clear to see that there also exists a minimum at $g=\sqrt{\gamma(\gamma+\kappa)+4\eta^2}/2$ for two atoms system. Compared with the single atom case, the value of $g^{(2)}(0)$ at $g=\sqrt{\gamma(\gamma+\kappa)+4\eta^2}/2$ decreases significantly [see red curve], leading to an improvement of the photon blockade phenomenon. The physical mechanism of this PB improvement attributes to the enhanced destructive interference effect resulting from the collective coupling enhancement [see Fig.~\ref{fig:fig3}(b)]. 
%Therefore, the quantum interference induced PB can also be improved via this collective coupling enhancement. 
%$g^{(2)}(0)\approx3.3\times10^{-4}$ shows a strong antibunching effect at an analytical predicted value of $g\approx0.43\kappa$ with $\gamma=\kappa$, and $g^{(2)}(0)\approx3.7\times10^{-4}$ at $g\approx0.71$ with $\gamma=\kappa/2$. It is noted that the optimal condition is same to the single atom case, while the collective behavior enhances the coupling strength between the $|+,1\rangle\leftrightarrow|gg,2\rangle$, leading resonant transition pathways (b) will be enhanced, and the prefect quantum destructive interference happens responsible for strong antibunching.

\vskip 5pt
{\it Constructive quantum interference leading to no photon blockade in atom-driven scheme.} - Let's consider that two atoms are driven by the coherent field. As shown in Fig.~\ref{fig:fig3}(c), there exist two different transition pathways for two-photon excitations, which is opposite to the single atom case. Using the same basis states, these two transition pathways can be represented by $|gg,0\rangle\overset{\sqrt{2}\eta}{\rightarrow}|+,0\rangle\overset{\sqrt{2}g}{\rightarrow}|gg,1\rangle \overset{\sqrt{2}\eta}{\rightarrow}|+,1\rangle\overset{2g}{\rightarrow}|gg,2\rangle$ and  $|+,0\rangle\overset{\sqrt{2}\eta}{\rightarrow}|ee,0\rangle \overset{\sqrt{2}g}{\rightarrow}|+,1\rangle\overset{2g}{\rightarrow}|gg,2\rangle$, respectively. To realize the PB, the excitation of the state $|+,1\rangle$ is required to be inhibited by the destructive interference so that the state $|gg,2\rangle$ remains unexcited. However, these two transitions are symmetric and indistinguishable so that the interference between these two transitions are constructive~\cite{PhysRevLett.93.093002}. Thus, the excitation of the state $|+,1\rangle$ is allowed in this case, and can be significantly enhanced via the constructive interference. As shown in Fig.~\ref{fig:fig4}(b), the value of $g^{(2)}(0)$ becomes much larger than that in single atom-driven case. 
%However, the $g^{(2)}(0)$ of the cavity transmission is strikingly different for cavity-driven case, as shown in Fig.\ref{fig:fig4}(b). The reason is these two excitation paths are indistinguishable, and the system can not exhibit the phenomenon of quantum destructive interference. On the other hand, the time-ordered pathways for two pairings can be suppressed the two-photon absorption induced by the quantum destructive interference. Thus, for the two atoms atom-driven QED system, it is possible to use special detunings in modifying the excite paths properties and then UPB can be achieved successfully.  

\section{Two atoms-driven cavity QED system with $\omega_a\neq\omega_c$}
\begin{figure}[htbp]
	\centering
	\includegraphics[width=\linewidth]{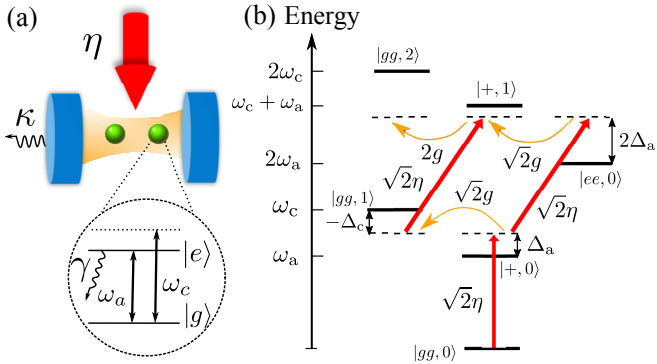}
	\caption{(Color online) (a) Sketch of two atoms-driven cavity QED system, where two atoms are driven by a coherent field $\eta$ and the atomic resonant frequency $\omega_a\neq\omega_c$. (b) The corresponding transition pathways show the destructive interference process.}~\label{fig:fig5}
\end{figure}
Here, we also study the two atoms-driven case but $\omega_a\neq\omega_c$, i.e., the atomic and cavity resonant frequency are different [see Fig.~\ref{fig:fig5}(a)]. In this system, the amplitude equations are given by 
%the case of Before introducing the full treatment of the open quantum system, we first derive the analytical expressions for the system and find optimal conditions that maximize the anthibuching character of the output field. Considering there is an out-of-resonance coupling between the atoms and the cavity in the CQED system, and the atoms simultaneously driven coherently, as shown in Fig.\ref{fig:fig5}(a). In the weak driving limit, the state of the system can be approximated in the two-photon manifold as
%
%\begin{eqnarray}\label{eq:bare state}
%&&|\Psi\rangle=C_{gg0}|gg,0\rangle+C_{gg1}|gg,1\rangle+C_{ge0}|ge,0\rangle+C_{eg0}|eg,0\rangle\nonumber\\
%&&+C_{gg2}|gg,2\rangle+C_{ge1}|ge,1\rangle+C_{eg1}|eg,1\rangle+C_{ee0}|ee,0\rangle,
%\end{eqnarray}
%%
%where $|jk,n\rangle=|j\rangle_1|k\rangle_2|n\rangle$ is the bare state, $|j\rangle_1$ and $|k\rangle_2$  denote the atomic state, and $|n\rangle$ is the number states of photon for cavity mode, respectively. Substituting the wave function in Eq.(\ref{eq:bare state}) and the Hamiltonian Eq.(\ref{eq:H2}) plus the dampings of cavity and atoms into the Schr\"{o}dinger equation, the coefficients $C_{jkn}$ are satisfied to
%
\begin{subequations}\label{eq:c2}
\begin{align}
i\dot{C}_{gg,1}=&\sqrt{2}gC_{+,0}-\left(i\frac{\kappa }{2}+\Delta _c\right)C_{gg,1}+\sqrt{2}\eta C_{+,1},\\
i\dot{C}_{gg,2}=&2gC_{+,1}+(-2 \Delta _c-i \kappa)C_{gg,2},\\
i\dot{C}_{+,0}=&\sqrt{2}\eta C_{gg,0}-\left(\Delta _a+i\frac{\gamma }{2}\right)C_{+,0}+\sqrt{2}gC_{gg,1}\nonumber\\
&+\sqrt{2}\eta C_{ee,0},\\
i\dot{C}_{+,1}=&\sqrt{2}\eta C_{gg,1}-\left(\Delta _a+\Delta _c+i\frac{\gamma+\kappa}{2}\right)C_{+,1}\nonumber\\
&+2 gC_{gg,2}+\sqrt{2}gC_{ee,0},\\
i\dot{C}_{ee,0}=&\sqrt{2}\eta C_{+,0}-(2 \Delta _a+i \gamma)C_{ee,0} +\sqrt{2}gC_{+,1}.
\end{align}
\end{subequations}
Under the steady state approximation and assuming $C_{gg,0}\gg\{C_{gg,1}, C_{+,0}\}\gg\{C_{gg,2},C_{+,1},C_{ee,0}\}$ and $C_{gg,0}\simeq1$, one can obtain 
\begin{subequations}\label{eq:cgg2-analysis}
\begin{align}
&C_{gg,2}=\frac{16 \sqrt{2} g^2 \eta ^2 [-2 i(2 \Delta _a+ \Delta _c)+2 \gamma +\kappa ]}{X(Y-2i\Delta_aZ)},\\
&C_{gg,1}=-\frac{8 g \eta }{8 g^2-\Delta _a \left(4 \Delta _c+2 i \kappa \right)-2 i \gamma  \Delta _c+\gamma  \kappa},\\
&C_{+,1}=\dfrac{8 \sqrt{2} g \eta ^2 \left(\kappa -2 i \Delta _c\right) [2(2 \Delta _a+\Delta _c)+i (2 \gamma +\kappa )]}{X(Y-2i\Delta_aZ)},
\end{align}
\end{subequations}
where $X=\Delta _a \left(-4 \Delta _c-2 i \kappa \right)-2 i \gamma  \Delta _c+\gamma  \kappa +8 g^2$ and $Y=-4 \Delta _a^2 \left(\kappa -2 i \Delta _c\right)+\gamma ^2 \kappa -4 \gamma  \Delta _c^2-2 i \Delta _c \left(\gamma ^2+2 \gamma  \kappa +4 g^2\right)+\gamma  \kappa ^2+8 \gamma  g^2+4 g^2 \kappa$ and $Z=-4 i \Delta _c (\gamma +\kappa )-4 \Delta _c^2+2 \gamma  \kappa +8 g^2+\kappa ^2$. Assuming $\{\Delta_a, \Delta_c\}\gg \{\gamma, \kappa\}$, the equal-time
second-order correlation can be expressed as
\begin{align}\label{eq:g2-analysis}
&g^{(2)}(0)\simeq\dfrac{2|C_{gg,2}|^2}{|C_{gg,1}|^4}\simeq\dfrac{(2\Delta_a+\Delta _c)^2(8 g^2-4\Delta_a\Delta _c)^2}{D},
\end{align}
where $D=[-8 \Delta _a \Delta _c (\gamma +\kappa )-4 \kappa  \Delta _a^2+\gamma ^2 \kappa -4 \gamma  \Delta _c^2+\gamma  \kappa ^2+8 \gamma  g^2+4 g^2 \kappa]^2+[8 \Delta _a^2 \Delta _c-2 \Delta _a (-4 \Delta _c^2+2 \gamma  \kappa +8 g^2+\kappa ^2)-2 \Delta _c (\gamma ^2+2 \gamma  \kappa +4 g^2)]^2$. Obviously, there exist two conditions to achieve $g^{(2)}(0)\rightarrow0$, yielding an  output of antibunched photons.

The first condition is $\Delta_a\Delta_c=2g^2$, corresponding to the traditional PB. In particular, this condition can be reduced to $\Delta_a=\pm\sqrt{2}g$ if one assume $\Delta_c=\Delta_a$, which is widely known as the condition for vacuum Rabi splitting and photon blockade phenomenon in strong coupling regime~\cite{zhu2017collective}. It is worth to point out that the atom-cavity coupling strength must be strong enough to observe the PB at the frequencies $\Delta_a\Delta_c=2g^2$, i.e. the strong coupling is critically important.

The second condition is $\Delta_c=-2\Delta_a$. Surprisingly, we find that it is independent to the atom-cavity coupling strength. 
%One of the condition for $g^{(2)}(0)\rightarrow0$ from Eq.(\ref{eq:g2-analysis}) is
%%
%\begin{equation}\label{eq:optimal1}
%\Delta_c=-2\Delta_a,
%\end{equation}
%
%where the intrinsic properties of the system($g, \kappa, \gamma$) do not appear, and the constraint equating which descried in Eq.(\ref{eq:one atom optimal g20}) is relaxed. It is should be noted that this optimal condition cannot be explained by the anharmonicity of the ladder of energy eigenstates, where the coupling strength $g$ should be larger than the loss rates in the system. This optimal condition for the two atoms coupled to the single cavity with atom-driven that we will focus on.
%
To understand the physical mechanism of this condition, we examine these two transition pathways for two-photon excitations again. As shown in Fig.~\ref{fig:fig5}(b), transitions (I) $|+,0\rangle\rightarrow|gg,1\rangle\rightarrow|+,1\rangle\rightarrow|gg,2\rangle$ and (II) $|+,0\rangle\rightarrow|ee,0\rangle\rightarrow|+,1\rangle\rightarrow|gg,2\rangle$ are distinct as opposite to the case of $\omega_a=\omega_c$. Thus, the destructive interference will take place if this specific condition is satisfied. In general, the occupying probability of state $|+,1\rangle$ obeys the second-order Fermi golden rule~\cite{PhysRevLett.93.093002,PhysRevA.92.023825}, yielding
\begin{equation}\label{eq:c}
|C_{+,1}|^2\propto\frac{\pi}{\hbar}\left|\frac{1}{\omega_a+\omega_c-2\omega_p}+\frac{1}{\omega_a-\omega_p}\right|^2.
\end{equation}
Clearly, the destructive interference will result in $|C_{+,1}|^2\rightarrow0$ if $-(\omega_a+\omega_c-2\omega_p)=\omega_a-\omega_p$ is satisfied. As a result, there is no population in the state $|gg,2\rangle$ and the second-order correlation function $g^{(2)}(0)\rightarrow0$. This condition implies that the quantum interference induced PB can be implemented in two atoms-driven cavity QED system if $\omega_a\neq\omega_c$. It is worth to point out that the physical mechanism of this destructive interference between even-photon transitions is different from that extensively studied in literature, where the interference occurs via the odd-photon transitions.
%the links between the bare states imposed by Eq.(\ref{eq:bare state}) and the corresponding transition paths are plot in Fig.\ref{fig:fig5}(b). We see that when $\Delta_a\neq\Delta_c\neq0$, the order of photons  exchanged($|+,0\rangle\leftrightarrow|g,1\rangle$) or absorbed($|+,1\rangle\rightarrow|ee,0\rangle$) first is uncertain. In this case, these two excitation paths are distinct and clearly distinguishable. According to the situation shown in Fig.\ref{fig:fig5}(b), 

%If $\omega_p-\omega_c$ and $\omega_p-\omega_a$ are distinct and nonzero, Eq.(\ref{eq:c}) leads to the  possibility of a minimum in state $|+,1\rangle$ corresponding to the $-(\omega_a+\omega_c-2\omega_p)=\omega_a-\omega_p$. Thus, the destructive interference of time-ordered excitation pathways occurs when the $|gg,1\rangle$ and $|ee,0\rangle$ contributions to $|+,1\rangle$ exactly cancel each other, leading to the strong antibuching.  The conclusion is consistent with the analytical result in Eq.(\ref{eq:optimal1}).

%
\begin{figure}[htbp]
	\centering
	\includegraphics[width=\linewidth]{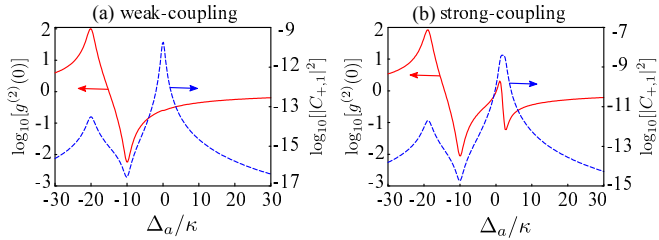}
	\caption{(Color online) Plots of the equal-time second-order correlation function $g^{(2)}(0)$ (red solid curves) and the probability for detecting state $|C_{+,1}|^2$ (blue dashed curves) as a function of the normalized detuning $\Delta_a/\kappa$. Here, we choose $\gamma=\kappa$, $\Delta_c=20\kappa$ and $\eta=0.01\kappa$, respectively. The coupling strength is given by (a) $g=0.5\kappa$ and (b) $g=5\kappa$, respectively.}~\label{fig:fig6}
\end{figure}
To verify these analytical results, we carry out numerical simulation by solving the master equation with system parameters $\Delta_c=20\kappa$ and $\gamma=\kappa$. For weak atom-cavity coupling strength, e.g., $g=0.5\kappa$, only a single minimum at the frequency $\Delta_a=-\Delta_c/2$ can be observed in the second-order correlation function [see Fig.~\ref{fig:fig6}(a), red curve]. As shown in Fig.~\ref{fig:fig6}(a), the value of $g^{(2)}(0)$ is smaller than $10^{-2}$ due to the destructive interference, resulting in a strong PB effect. Simultaneously, the value of $|C_{+,1}|^2$ (blue curve) reaches its minimum, leading to $|C_{gg,2}|^2\rightarrow0$. For strong atom-cavity coupling strength (e.g., $g=5\kappa$), however, there exist two minimums in the second-order correlation function, corresponding to the frequencies $\Delta_a=-\Delta_c/2$ and $\Delta_a=2g^2/\Delta_c$, respectively [see Fig.~\ref{fig:fig6}(b), red curve]. The values for these two minimums are both less than unity. The left one is resulted from the destructive interference since $|C_{gg,2}|^2$ also reaches its minimum [see the blue curve], while the right one attributes to the anharmonic energy splitting. It is clear to see that the value of $g^{(2)}(0)$ at $\Delta_a=-\Delta_c/2$ is much smaller than that at $\Delta_a=2g^2/\Delta_c$.
\begin{figure}[htbp]
	\centering
	\includegraphics[width=\linewidth]{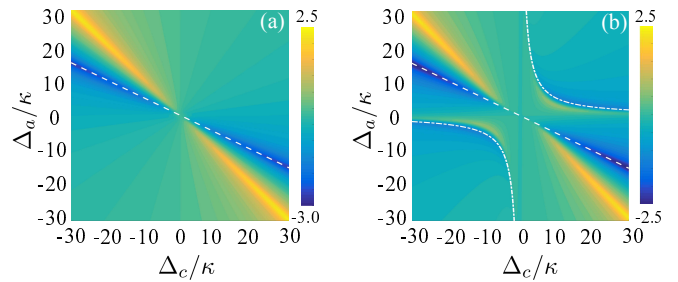}
	\caption{(Color online) Equal-time second-order correlation function  $g^{(2)}(0)$ versus the cavity detuning $\Delta_c$ and atomic detuning $\Delta_a$, respectively. Here, the coupling strength is chosen as (a) $g=0.5\kappa$ and (b) $g=5\kappa$, respectively. The white dashed curves represent the traditional PB condition ($\Delta_a\Delta_c=2g^2$) and the white dash dotted curves denote the condition of the interference induced PB ($\Delta_c=-2\Delta_a$). The other parameters are same as those used in Fig.~\ref{fig:fig6}.}~\label{fig:fig7}
\end{figure}
In Fig.~\ref{fig:fig7}(a), we plot the second-order correlation function $g^{(2)}(0)$ as functions of the detunings $\Delta_a$ and $\Delta_c$, respectively. In panels (a) and (b), we chose the atom-cavity coupling strength $g=0.5\kappa$ and $g=5\kappa$, respectively. Other system parameters are the same as those used in Fig.~(6). The dashed and dash-dotted curves indicate the analytical expressions $\Delta_a=-\Delta_c/2$ and $\Delta_a\Delta_c=2g^2$, respectively. It is clear to see that the second order correlation function at $\Delta_a=-\Delta_c/2$ is always smaller than unity in both weak and strong coupling regimes. However, the correlation function at $\Delta_a\Delta_c=2g^2$ is smaller than unity only when the system enters into the strong coupling regime, which matches the analytical results very well.
%To prove that Eq.(\ref{eq:optimal1}) and Eq.(\ref{eq:optimal2}) resulting in a strong antibunching effect, we show a full $g^{(2)}(0)$ map versus $\Delta_a$ and $\Delta_c$ for $g=0.5\kappa$(panel (a)) and $g=5\kappa$(panel (b)) by numerically In Fig.\ref{fig:fig7}. The map reveals the occurrence of the UPB(dashed white line) induced by the quantum interference mechanism both in weak- and strong-coupling regime. While PB(green dasehed line) caused by anharmonicity of energy spectrum only occurs in strong-coupling regime due to the nonlinear requirement. Hence, numerical results are consistent with analytical calculations.

%
\begin{figure}[htbp]
	\centering
	\includegraphics[width=\linewidth]{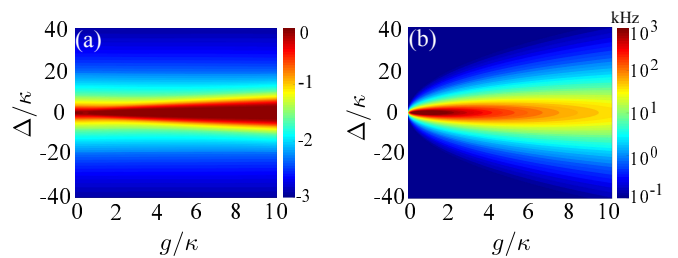}
	\caption{(Color online) Plots of the equal-time second-order correlation function $g^{(2)}(0)$ [panel (a)] and the counting rate [panel (b)] as functions of the detuning $\Delta\equiv\Delta_a=-\Delta_c/2$ and the coupling strength $g$, respectively. The system parameters are chosen as $\kappa/2\pi=2.8$ MHz, $\gamma/2\pi=3.0$ MHz, $\eta/2\pi=1.4$ MHz, which are the same as those given in Ref.~\cite{neuzner2016interference}.}~\label{fig:fig8}
\end{figure}
Finally, we discuss the influence of the atom-cavity coupling strength on the interference induced PB in this two atoms cavity QED system by setting $\Delta\equiv\Delta_a=-\Delta_c/2$. In Fig.~\ref{fig:fig8}(a), the second-order correlation function $g^{(2)}(0)$ is plotted as functions of the normalized atom-cavity coupling strength $g/\kappa$ and the normalized detuning $\Delta/\kappa$, respectively. Here, we choose a set of experimental parameters as $\kappa/2\pi=2.8$ MHz, $\gamma/2\pi=3.0$ MHz, $\eta/2\pi=1.4$ MHz~\cite{neuzner2016interference}. We notice that $g^{(2)}(0)$ decreases quickly as the detuning $\Delta$ increases, but changes slightly as the increase of the atom-cavity coupling strength $g$. Contrary to the second order correlation function, the counting rate of cavity photons [see panel (b)] grows significantly as the atom-cavity coupling strength $g$ increases. Therefore, detectable photons with strong antibunching behavior can be accomplished under strong coupling regime if a specific detuning $\Delta$ is chosen. 
%This shows that under the optimal conditions, the UPB is not sensitive to the coupling strength. After considering the parameters in the experiments\cite{hamsen2017two}, we take the $\kappa=2\pi\times2$MHz and $\gamma=2\pi\times3$MHz for numerical calculations, the other adjustable parameter is selected as $\Delta_c=30\kappa$, $\Delta_a=-15\kappa$ and $\eta=0.4\kappa$, $g^{(2)}(0)$(red solid line) and the spectra(blue dashed line, $\kappa\langle a^\dagger a\rangle$) as a function of the coupling strength $g$ is plotted in Fig.\ref{fig:fig8}(b). It can be seen that $g^{(2)}(0)\approx0.01$ basically does not change with increase of $g$, but the mean photon number in the cavity increases significantly.

\section{Conclusion}
In summary, we have theoretically investigated the nature of the interference induced photon blockade in cavity QED with one and two atoms. In a single atom cavity QED system, we show that the interference induced photon blockade can only be observed when the external field drives the cavity directly. In the atom driven scheme, there exists a single transition pathway for the two-photon excitation so that the interference induced photon blockade cannot be observed. In two atoms cavity QED system, the quantum interference induced photon blockade still exists in cavity driven scheme. In the atom-driven case, we show that there exist two transition pathways for the two-photon excitation as opposite to the single atom case. If the atomic and cavity resonant frequencies are the same, these two transition pathways are indistinct and lead to constructive interference, which is harmful to the photon blockade effect. However, if the atomic and cavity resonant frequencies are different, these two transition pathways become distinct and lead to a new kind of photon blockade effect based on the even photon destructive interference. Moreover, we show that the condition for this novel interference induced photon blockade is independent to the atom-cavity coupling strength, which provides us the possibility for observing large photon number with strong antibunching behavior in the strong coupling regime. 
%the optimal antibunching character of the atoms coupled to a single optical cavity under different system driving mode induced by quantum interference is investigated numerically and theoretically. It is demonstrated that the antibunching character can be optimized by tuning of the intrinsic system parameters for cavity-driven case, while only depend on the relationship between atom-drive and cavity-drive detunings when the atoms are driven. Additionally, the independence of coupling strength and the cavity decay rate suggest that the optimization mechanism is of good robustness. Nowdays, UPB has been observed in two coupled superconducting resonators system\cite{PhysRevLett.121.043602} and quantum dot cavity QED system\cite{PhysRevLett.121.043601}.  We hope that the proposed scheme will provide us with a way to control the PB exactly in atom CQED system.

\vskip 5pt
\begin{acknowledgments}
CJZ and YPY thank the support of the National Key Basic Research Special Foundation (Grant No.2016YFA0302800); the Shanghai Science and Technology Committee (Grants No.18JC1410900); the National Nature Science Foundation (Grant No.11774262). KH thanks the support of the Natural Science Foundation of Anhui Province (Grant No.1608085QA23). Natural Science Foundation of Anhui Provincial Education Department (Grant No.KJ2018JD20). GSA thanks the support of Air Force Office of  scientific Research (Award No. FA-9550-18-1-0141).
\end{acknowledgments}

\bibliography{refs}

\end{document}